\documentclass[runningheads]{llncs}
\usepackage[T1]{fontenc}
\usepackage[compatibility=false]{caption}

\usepackage{graphicx}
\usepackage{longtable}   
\usepackage{booktabs}    
\usepackage{caption}     
\usepackage{array}       
\usepackage{tcolorbox}
\usepackage{subcaption} 
\usepackage{url}
\usepackage{amsmath}
\usepackage{amssymb}

\usepackage{microtype}

\begin{document}
\title{Early Prediction of Cross-Service Implementation Scope: Evidence and Limits from Multiple Microservice Projects}
\titlerunning{Early Prediction of Cross-Service Implementation Scope}
%
\author{
Zhoumaocuo\inst{1}
\and
Xiaozhou Li\inst{2}\orcidID{0000-0002-3767-2527}
\thanks{Corresponding author.}
}
\authorrunning{Zhoumaocuo and Li}
%
\institute{
Virginia Polytechnic Institute and State University,
Blacksburg, VA, USA\\
\email{zmaocuo0219@vt.edu}
\and
Xi'an Jiaotong-Liverpool University,
Suzhou, China \\
\email{xiaozhou.li@xjtlu.edu.cn}
}
\maketitle              
\begin{abstract}

Microservice architectures aim to localize change, yet feature requests may require implementation across service boundaries. We investigate how well such downstream scope can be anticipated before implementation begins. Using issue--pull-request--commit traceability and implementation evidence corrected through targeted human adjudication, we reconstruct 866 repository-traceable feature requests from 13 open-source microservice projects and map their implementations to validated service boundaries. Cross-service changes occur in 101 requests (11.7\%) and exhibit broader implementation footprints, particularly higher code churn, but no clear difference in lead time. Pre-implementation evidence provides limited cross-project predictive signal: the strongest leave-one-system-out model reaches a project-macro Macro-F1 of 0.484, but its advantage over the majority baseline is not significant after Holm correction and is sensitive to small evaluation folds. A local LLM offers no reliable improvement over classical models. Downstream workload is substantially harder to estimate, as learned models do not consistently outperform training-fold median baselines. A simple service-mention heuristic also performs comparably to learned scope triage. These findings show that early feature-request evidence is more useful for lightweight cross-service scope triage than for automated scope determination or precise workload estimation.

\end{abstract}
\section{Introduction}
\label{sec:introduction}

Microservice architectures aim to localize change through independently deployable services, yet implementation can still propagate across service boundaries through technical dependencies and evolutionary relationships \cite{amoroso2024understanding,d2023microservice,cerny2025towards}. This creates an early planning question for maintainers: \emph{can the eventual architectural reach of a feature request be anticipated before implementation begins?}

Cross-service work is also an organizational phenomenon. Developers frequently contribute to multiple services \cite{amoroso2023one,li2025exploring}, while Organizational Coupling (OC) captures such activity \cite{li2023evaluating} and has been linked to developer roles, organizational cohesion, and architectural degradation \cite{li2026keydeveloperroles,li2026organizational,mani2026organizational}. Together with logical-coupling and change-impact studies \cite{d2023microservice,cerny2025towards}, this work explains cross-service evolution mainly after development activity becomes observable.

The earlier decision point at feature intake remains less understood. Issue trackers provide request text, labels, reporter history, repository context, and early discussion that may support prediction. Although issue text has been used for effort estimation \cite{choetkiertikul2018deep}, anticipating \emph{architectural reach} is distinct: estimating whether implementation will cross service boundaries is not equivalent to estimating how much downstream work will follow.

We distinguish \emph{implementation scope}, the number of microservices touched by a realized change, from \emph{implementation workload}, represented by commits, developers, code churn, and lead time. Scope is not a direct proxy for effort, nor is cross-service implementation inherently harmful; legitimate features, dependency changes, and coordinated maintenance may span several services. Our goal is therefore to anticipate broader architectural reach for planning and review.

We study this through multi-project repository mining, reconstructing and human-validating implementation evidence from issue--pull-request--commit traceability and mapping changed files to validated service boundaries.
Predictors are limited to pre-implementation evidence. We compare text-based and structured classical models, their combination, and a local LLM, using leave-one-system-out validation as the primary cross-project evaluation. This design addresses the following research questions: 


\smallskip
\noindent\textbf{RQ1.} \emph{How does implementation scope vary across microservice feature requests?}

\smallskip
\noindent\textbf{RQ2.} \emph{How accurately can pre-implementation evidence predict cross-service implementation?}

\smallskip
\noindent\textbf{RQ3.} \emph{How accurately can downstream workload be estimated before implementation?}

\smallskip
\noindent\textbf{RQ4.} \emph{Which early information supports scope prediction and practical triage?}

Our cohort contains 866 repository-traceable feature requests with targeted human corrections from 13 projects, including 101 (11.7\%) cross-service implementations. Cross-service requests show broader realized work, particularly code churn, but remain difficult to predict: the strongest primary LOSO model reaches Macro-F1=0.484, although its advantage over the majority baseline is not significant after Holm correction and is sensitive to small test folds. The local LLM offers no reliable improvement. Workload estimation is weaker still, with learned models failing to consistently outperform training-fold median baselines. Overall, early evidence is more informative about where implementation may spread than how much work will follow.

This paper contributes: (1) a multi-project characterization of implementation scope and downstream workload using human-adjudicated implementation evidence; (2) a leakage-controlled evaluation of classical and local-LLM approaches that establishes the limits of early scope and workload prediction; and (3) an analysis of pre-implementation information for lightweight scope triage, including comparisons with simple heuristics. The results support prioritizing requests for earlier architectural review rather than automated implementation planning.

\section{Background and Related Work}
\label{sec:relatedwork}

Microservice architecture decomposes a system into small, independently deployable services, aiming to support localized change and autonomous evolution \cite{dragoni2017microservices}. This idea builds on long-standing modularity principles, where good decomposition should hide design decisions and reduce unnecessary coupling \cite{parnas1972}. In practice, however, microservice systems still contain technical dependencies that may cause changes to propagate across service boundaries~\cite{abdelfattah2023microservice,abdelfattah2025multivocal}. Prior work has therefore studied microservice coupling, logical co-change, maintainability, and change impact after implementation evidence becomes available~\cite{zhong2023measuring,amoroso2024understanding,cerny2025towards,cerny2025analyzing}.

A complementary research stream examines the socio-technical side of microservice evolution. Conway's Law suggests that software structure and organizational structure co-evolve~\cite{conway1968committees}, while socio-technical congruence links coordination needs with development activity~\cite{cataldo2008socio}. In microservice projects, empirical studies show that developers often contribute to multiple services rather than following a strict one-service ownership model~\cite{amoroso2023one,li2025exploring}. Organizational Coupling further quantifies cross-service developer activity~\cite{li2023evaluating,li2025exploring}, and recent work relates such activity to collaboration structures, key developer roles, organizational cohesion, and architecture degradation~\cite{li2023analyzing,li2026keydeveloperroles,li2026organizational,mani2026organizational}.

These studies explain how services and developers become coupled during system evolution, but they mainly rely on evidence observed after changes have occurred. Feature requests raise an earlier question: can maintainers infer implementation scope before the implementation begins? Issue trackers contain rich pre-implementation information, including requirements, labels, and discussion, and prior work shows that microservice issues capture diverse maintenance and architectural concerns~\cite{waseem2021}. Yet it remains unclear whether such early evidence can distinguish requests that will remain within one service from those that will require cross-service implementation, or whether it can estimate downstream workload. This paper addresses that gap by studying pre-implementation prediction of feature-request scope and workload across multiple open-source microservice projects.

\section{Study Design}
\label{sec:studydesign}

We conduct a multi-system repository-mining study to assess whether pre-implementation evidence can anticipate (1) the realized service scope of a feature request and (2) its downstream repository-observable workload. Request-time information forms the predictors, while linked implementation commits and service-mapped file changes provide the outcomes. 
Classical models and a locally executed LLM are evaluated using adjudication-corrected implementation outcomes, without manual prediction assessment.


\subsection{Subject Systems and Data}
\label{sec:subjects}

The candidate corpus consists of 15 open-source microservice systems previously selected from a manually validated microservice dataset \cite{amoroso2024dataset}. The systems vary in service count, age, repository activity, issue volume, and developer population. For each project, four aligned sources---issues, issue timelines, commits, and developer mappings---provide request information, repository-native traceability, changed files, and validated file-to-service mappings. Developer aliases are normalized and automated accounts are removed consistently.

Validated service directories determine implementation scope. Repository-level files outside these directories contribute to code churn but do not create additional services. Projects are retained only when the common feature identification, traceability, temporal, and service-mapping criteria below yield eligible requests; this decision is independent of prediction performance. Of the 15 candidate systems, 13 contribute eligible requests to the final cohort. The distribution is highly uneven: Camunda contributes 720 of 866 requests (83.1\%), while eight projects contain six or fewer requests. Project-level retention counts and exclusion reasons are documented in the replication package.

\subsection{Feature Requests and Implementation Ground Truth}
\label{sec:cohort}

The unit of analysis is an implemented non-PR GitHub issue. Feature requests are identified from normalized repository-native labels containing \emph{feature}, \emph{enhancement}, or \emph{improvement}; normalization rules and label vocabulary are included in the replication package. Pull requests are excluded as prediction units using PR-specific timeline evidence but remain usable as implementation evidence.

\noindent\textit{Prediction checkpoints.}
For request $i$, let $t_i^{\mathrm{create}}$ denote its creation time and $t_i^{\mathrm{label}}$ the first qualifying feature-label event. The primary prediction checkpoint is

\begin{equation}
    t_i^0 = \max(t_i^{\mathrm{create}},
                 t_i^{\mathrm{label}}).
\end{equation}

Only requests whose first traceable implementation occurs after $t_i^0$ are retained, preventing retrospective use of feature labels. RQ4 additionally evaluates $t_i^{24}=t_i^0+24$ hours; requests whose implementation begins within this interval are excluded from that comparison.

\noindent\textit{Request-to-implementation traceability.}
Implementation commits are reconstructed from repository-native direct commit references and commits associated with cross-referenced PRs. Let $C_i^{\mathrm{dir}}$ denote directly referenced commits, $P_i$ the linked PRs, and $C_p$ the commits of PR $p$:

\begin{equation}
    C_i = C_i^{\mathrm{dir}} \cup
          \bigcup_{p\in P_i} C_p.
\end{equation}

Commit identifiers are de-duplicated, and constituent PR commits are preferred over redundant merge commits. PRs linked to multiple issues may introduce overlapping implementation evidence; suspicious issue--commit associations are examined individually, and unsupported links are removed. We avoid heuristic text-based matching; unlinked requests are excluded rather than treated as zero-effort implementations.

The cohort construction starts from 7,708 feature-labeled issues. We exclude 1,627 PRs, 495 open or invalid-$t^0$ requests, 3,890 without recoverable implementation traceability, 285 without implementation after $t^0$, and 523 without service-mapped implementation, yielding 888 initially eligible requests. Targeted human adjudication corrected 69 Camunda requests: 21 changed from cross-service to single-service, 26 received evidence-only corrections, and 22 became unobservable and were excluded. The final frozen cohort contains 866 requests from 13 projects, including 101 cross-service implementations; 695 satisfy the $t^{24}$ eligibility rule. An independent 50-case probability sample, reserved for validation rather than cohort correction, estimates a design-based residual labeling error of 3.0\% (95\% CI: 0--6.7\%). The study therefore concerns repository-traceable implementations rather than all feature requests.

For 41 adjudication disagreements, Claude (model 'claude-opus-5-5') generated draft reconciliation proposals based on the available implementation evidence. Both human reviewers examined the evidence and jointly approved the final decisions, recording their rationales. No AI-generated proposal was accepted automatically. The resulting ground-truth corrections therefore remained subject to human review and approval.

\subsection{Outcomes and Pre-Implementation Predictors}
\label{sec:variables}

For each request $i$, the implementation service set $S_i$ is the union of all validated services touched by commits in $C_i$. The primary target is

\begin{equation}
Y_i^{CS}=
\begin{cases}
0, & |S_i|=1,\\
1, & |S_i|\geq2,
\end{cases}
\label{eq:scope}
\end{equation}

distinguishing single-service from cross-service implementation. This label captures implementation breadth rather than architectural quality: a cross-service change may represent legitimate feature work, dependency evolution, or coordinated maintenance.

RQ3 considers four repository-observable workload outcomes: linked commit count, unique normalized implementation developers, code churn (additions plus deletions), and lead time from $t_i^0$ to the latest linked implementation commit. These measures characterize realized implementation footprint rather than person-hours or monetary cost.


Predictor values use events timestamped no later than the relevant checkpoint, except that service mentions are matched against the retrospectively validated project service vocabulary. Table~\ref{tab:features} summarizes the feature families.

\begin{table}[!ht]
\centering
\caption{Pre-implementation predictor families.}
\label{tab:features}
\scriptsize
\begin{tabular}{p{0.20\linewidth}p{0.75\linewidth}}
\toprule
\textbf{Family} & \textbf{Predictors} \\
\midrule
Request semantics &
Title/body text, length, service/component mentions. \\

Request metadata &
Labels, author association, label count, explicit component/service. \\

Reporter history &
Prior requests, commits, services, and historical request scope. \\

Repository context &
30-/90-day activity, active developers/services, cross-service ratio. \\

Early triage &
24-h comments, participants, labels, discussion length, service mentions. \\
\bottomrule
\end{tabular}
\end{table}

These families capture complementary potential indicators of implementation scope: request semantics and metadata may identify affected components, reporter history reflects previous implementation patterns, and repository context captures recent development activity. The 24-hour features test whether early discussion provides additional information beyond request-time evidence.

Title and body text are represented with TF--IDF and reduced using truncated SVD fitted on training data only. Historical and repository-context features use only pre-checkpoint activity; missing history for new contributors remains missing, and post-checkpoint aggregates are excluded. After ground-truth adjudication, historical request-outcome features were recomputed from corrected earlier requests, affecting 300 Camunda requests without introducing post-checkpoint information. The complete predictor names, definitions, and checkpoint-availability rules are provided in the replication package.

\subsection{Prediction Models}
\label{sec:models}

For RQ2, we compare a majority baseline, TF--IDF logistic regression, structured XGBoost using metadata and historical/context features, and combined XGBoost incorporating the reduced text representation. The majority classifier provides a baseline under class imbalance, logistic regression offers a simple sparse-text comparator, and XGBoost tests nonlinear relationships among structured features and potential complementary information from text. Class weights are estimated within each training fold. These four configurations form the primary classical comparison.

We additionally evaluate the locally executed \texttt{gpt-oss:20b} model using only checkpoint-available text, labels, and the same historical context. Direct repository identifiers are removed, the inference configuration is fixed, and invalid outputs count as prediction errors. Its one-, two-, and three-or-more-service outputs are merged for the primary binary task. 
This provides a zero-shot comparison without task-specific training. The corrected-cohort rerun used the installed model build identified by digest prefix \texttt{9ba9cc2b}, as the originally submitted build was unavailable. Prompt construction, parsing, and inference settings were unchanged, and predictions were regenerated with a fresh cache.

For RQ3, a training-fold median baseline is compared with ridge and XGBoost regression on $\log(1+y)$-transformed outcomes, with training-only model selection. The median provides a distribution-based baseline, while ridge and XGBoost represent regularized linear and nonlinear alternatives, respectively.

\subsection{Evaluation Protocol}
\label{sec:evaluation}

\textit{RQ1: implementation footprint.} We report project-level and pooled scope/workload distributions using medians and IQRs, and compare single- and cross-service requests using Cliff's $\delta$ with bootstrap confidence intervals. Because pooled results can be influenced by unequal project sizes, secondary analyses also compute within-project effect sizes for projects containing both scope classes.

\textit{RQ2--RQ3: predictive evaluation.} Leave-one-system-out (LOSO) validation is primary: each eligible project is held out once, while preprocessing, feature transformation, class weighting, and model tuning are fitted using the remaining projects. Metrics are computed for each held-out project and macro-averaged so that large repositories do not dominate the evaluation. RQ2 uses Macro-F1 as the primary metric, complemented by balanced accuracy and cross-service precision, recall, and F1. Macro-F1 averages over both binary labels with \texttt{zero\_division=0}, including single-class test folds; balanced accuracy averages recall over classes present in each fold. RQ3 uses MAE, with RMSLE and Spearman's $\rho$ as complementary measures. As a temporal check, requests within eligible projects are ordered by $t_i^0$, with the earliest 70\% used for training and the latest 30\% for testing without reshuffling.

\textit{RQ4: information value and scope triage.} Feature-family ablations evaluate text, metadata, history/context, and their combinations; these analyses remain exploratory and do not redefine the primary RQ2 models. We compare $t_i^0$ and $t_i^{24}$ on the same requests eligible at both checkpoints. The 24-hour checkpoint is an early discussion opportunity, not a performance-tuned threshold, and implementation must not have begun by that time. Triage uses the all-family $t_i^0$ model to select the highest predicted cross-service-probability requests within each held-out project, with budget $\max(1,\lceil 0.2n\rceil)$ for $n$ requests. We report Precision@20\%, Recall@20\%, and Lift@20\% as project-macro values and pooled operational counts. As a secondary comparison, two fixed heuristics rank requests by the number of distinct service names mentioned in their $t_i^0$ text or the number of component/scope labels active at $t_i^0$. Both use the same budget; tied heuristic scores are evaluated by expected random tie-breaking.

\textit{Sensitivity analyses and leakage safeguards.} The frozen adjudication-corrected cohort of 866 requests from 13 projects is primary. We retain five secondary robustness checks: (1) repeat RQ1 and classical RQ2 without Camunda; (2) examine within-project RQ1 effects; (3) deduplicate exact implementation-signature groups; (4) restrict RQ2 to held-out projects with at least ten requests and both classes; and (5) compare direct-only, PR-only, and combined traceability mechanisms. We additionally evaluate seeds 42--45 without selecting the best-performing run; seed 42 remains primary.

Throughout, request-dependent predictors are frozen at $t_i^0$ or $t_i^24$; later issue activity and implementation evidence are reserved for ground-truth construction. Historical outcome features use adjudicated requests completed before the prediction checkpoint. Preprocessing and model selection occur within training folds. Bootstrap intervals quantify uncertainty, and paired comparisons use Wilcoxon signed-rank tests with Holm correction within comparison families.
\section{Results}
\label{sec:results}

We report project-macro results under leave-one-system-out (LOSO) validation unless stated otherwise. This gives each project equal weight when evaluating transfer to unseen systems. Our primary analysis uses the frozen cohort of 866 repository-traceable requests, with implementation evidence corrected through targeted human adjudication. The sensitivity analyses described in Section \ref{sec:evaluation} assess the stability of these findings.

\subsection{RQ1: Downstream Implementation Scope}
\label{sec:rq1-results}

The final cohort contains 866 feature requests from 13 projects: 765 (88.3\%) remained within one service, whereas 101 (11.7\%) required changes to multiple services. Of the latter, 43 touched two services and 58 touched three or more. Figure~1 shows that prevalence varied substantially across projects. Camunda alone contributes 720 requests, of which only 4.2\% are cross-service; consequently, the pooled 11.7\% should not be interpreted as a project-independent prevalence estimate. Removing Camunda increases the descriptive cross-service share to 48.6\%.

\begin{figure}[!ht]
    \centering
    \includegraphics[width=0.8\textwidth]{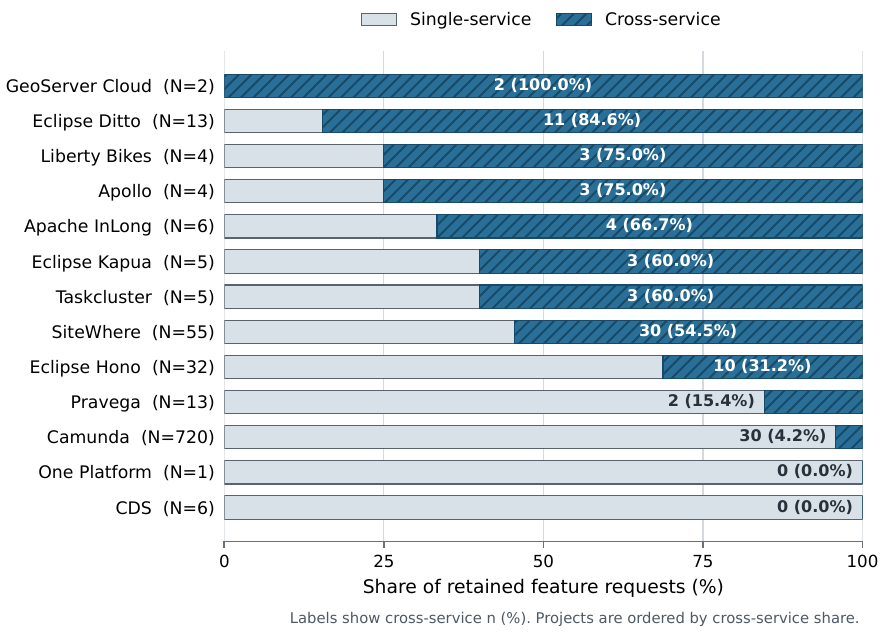}
    \caption{Project-level composition of single- and cross-service feature requests. Bars are normalized within projects; labels report the number and percentage of cross-service requests, with project sample sizes shown on the vertical axis.}
    \label{fig:rq1-scope}
\end{figure}

Cross-service requests nevertheless exhibited broader implementation footprints (Table \ref{tab:rq1-footprint}). Their median churn was 2,146 lines, compared with 521 for single-service requests, producing the largest pooled distributional difference ($\delta=0.484$). Smaller positive effects were observed for commits ($\delta=0.229$) and developers ($\delta=0.150$), whereas lead time showed no clear difference ($\delta=0.008$, 95\% CI [$-0.113$, $0.130$]). The churn pattern is not solely an artifact of the dominant project: after removing Camunda, its effect remains positive, as do those for commits and developers. Among the ten projects containing both scope classes, seven show positive churn effects. These associations do not establish that crossing service boundaries causes additional work.

\begin{table}[!ht]
\centering
\scriptsize
\caption{Implementation footprint by realized scope. Values are median (IQR); positive Cliff's $\delta$ indicates larger cross-service values.}
\label{tab:rq1-footprint}
\begin{tabular}{lrrrr}
\toprule
\textbf{Outcome} & \textbf{Single-service} & \textbf{Cross-service}
& $\boldsymbol{\delta}$ & \textbf{95\% CI} \\
\midrule

Commits & 2 (3) & 3 (7) & .229 & [.117, .344] \\
Developers & 1 (0) & 1 (1) & .150 & [.057, .244] \\
Churn (LOC) & 521 (1035) & 2146 (5135) & .484 & [.368, .588] \\
Lead time (d) & 18.0 (59.0) & 22.2 (61.5) & .008 & [$-.113$, .130] \\

\bottomrule
\end{tabular}
\end{table}

\textit{Answer to RQ1.} Pooled across the retained cohort, 11.7\% of requests are cross-service, but prevalence is strongly project-dependent. Such implementations tend to have broader repository-observable footprints, particularly code churn; a lead-time difference is not established.

\subsection{RQ2: Predicting Cross-Service Implementation}
\label{sec:rq2-results}

Table \ref{tab:rq2-prediction} reports the primary LOSO classification results. Among the original models, combined XGBoost obtains the highest project-macro Macro-F1, 0.484 [0.360, 0.604], compared with 0.309 for the majority baseline. Its improvement of 0.174 [0.012, 0.315], however, does not remain statistically significant after Holm correction ($p_{\mathrm{Holm}}=0.078$); differences from text logistic regression (+0.141) and structured XGBoost (+0.115) are also uncertain (both $p_{\mathrm{Holm}}=0.098$). Combined XGBoost has the highest classical cross-service F1 (0.514), compared with 0.408 for structured XGBoost, but no configuration demonstrates a robust advantage across evaluation conditions.

The local \texttt{gpt-oss:20b} model (build \texttt{9ba9cc2b}) produced valid terminal outputs for 776 of 866 requests (89.6\%) and reached Macro-F1 of 0.332 [0.190, 0.472] when invalid outputs were counted as errors. It did not improve over combined XGBoost ($\Delta=-0.151$, $p_{\mathrm{Holm}}=0.37$). A matched-prompt audit found 77.9\% agreement in fine-grained labels between model builds, indicating build-dependent results. Temporal validation over five eligible projects favors text logistic regression (0.499), followed by structured XGBoost (0.493) and combined XGBoost (0.458).

The sensitivity analyses reinforce the broader conclusion while showing that model ranking is unstable. Without Camunda, structured XGBoost is highest (Macro-F1=0.441); deduplication reduces combined XGBoost to 0.393, while the five adequately supported held-out projects yield 0.450. Across seeds 42--45, combined XGBoost ranges from 0.397 to 0.520. These changes, especially on small test folds, preclude a stable superiority claim.

\begin{table}[!ht]
\centering
\scriptsize
\setlength{\tabcolsep}{3pt}
\caption{LOSO prediction of single- versus cross-service implementation. Metrics are project-macro; 95\% CIs are shown for Macro-F1.}
\label{tab:rq2-prediction}
\begin{tabular}{lccccc}
\toprule
\textbf{Model} & \textbf{Macro-F1 [95\% CI]} & \textbf{Bal. Acc.}
& \textbf{CS P} & \textbf{CS R} & \textbf{CS F1} \\
\midrule
Majority & 0.309 [0.228, 0.391] & 0.538 & 0.000 & 0.000 & 0.000 \\
Text Logistic & 0.343 [0.284, 0.403] & 0.535 & 0.216 & 0.154 & 0.155 \\
Structured XGB & 0.369 [0.245, 0.494] & 0.450 & 0.394 & 0.509 & 0.408 \\
Combined XGB & 0.484 [0.360, 0.604] & 0.567 & 0.583 & 0.563 & 0.514 \\
Local LLM & 0.332 [0.190, 0.472] & 0.351 & 0.439 & 0.288 & 0.312 \\ 
\bottomrule
\end{tabular}
\end{table}

\textit{Answer to RQ2.} Pre-implementation evidence contains limited cross-project signal about whether implementation will cross service boundaries. Model ranking depends on the cohort and random seed, while neither additional model complexity nor the local LLM provides a consistent advantage.

\subsection{RQ3: Estimating Downstream Workload}
\label{sec:rq3-results}

The scope signal does not translate into reliable estimation of exact downstream workload. Table~4 shows that XGBoost does not reliably improve project-macro MAE over the training-fold median baseline for any target. It is significantly worse for developers ($p_{\mathrm{Holm}}=0.037$), while the higher error for commits falls short of significance after correction ($p_{\mathrm{Holm}}=0.056$); churn and lead-time differences are not distinguishable. Ridge regression is unstable, particularly for code churn. XGBoost rank correlations are weak or negative for all targets (from $-0.321$ to $0.076$).

\begin{table}[!ht]
\centering
\scriptsize
\caption{LOSO workload estimation from request-time evidence. Lower MAE is better; $\rho$ is the XGBoost Spearman correlation.}
\label{tab:rq3-workload}
\begin{tabular}{lrrrr}
\toprule
\textbf{Target} & \textbf{Median MAE} & \textbf{Ridge MAE}
& \textbf{XGB MAE} & \textbf{XGB $\rho$} \\
\midrule
Commits & 6.54 & 113 & 7.42 & .075 \\
Developers & .854 & 1.222 & 1.216 & $-.321$ \\
Churn & 4965 & $3.7\times10^{8}$ & 5047 & $-.021$ \\
Lead days & 166.1 & 165.6 & 165.9 & .076 \\
\bottomrule
\end{tabular}
\end{table}

Temporal validation produces the same general pattern. XGBoost improves over the median baseline only for developer count (MAE 1.39 versus 1.61), but not for commits, churn, or lead time.

\textit{Answer to RQ3.} Request-time evidence does not support reliable estimation of exact downstream workload. It is more informative about whether implementation may spread across service boundaries than about the number of commits, developers, changed lines, or elapsed days that will eventually follow.


\subsection{RQ4: Information Value and Practical Triage}
\label{sec:rq4-results}


In the exploratory feature-family analysis, text plus history/context yields the highest observed Macro-F1 (0.532 [0.447, 0.622]), compared with 0.501 when all request-time families are included. Other configurations range from 0.374 to 0.465, and removing any individual family causes no statistically robust deterioration (all $p_{\mathrm{Holm}}\geq0.12$). The ablation therefore does not identify a reliably dominant source of information and does not redefine the primary RQ2 model comparison.

The first 24 hours of issue activity provide no reliable improvement. Among the same 695 requests eligible at both checkpoints, Macro-F1 changes from 0.389 to 0.374 ($\Delta=-0.015$, 95\% CI [$-0.045$, $0.014$]).

Predicted cross-service probability from the all-family A6 configuration nevertheless offers modest ranking value. Selecting the highest-ranked 20\% independently within each held-out project captures 40 of 101 cross-service implementations. Project-macro Precision@20\%, Recall@20\%, and Lift@20\% are 0.572, 0.364, and 1.53, respectively. The pooled selected sets contain 178 requests, of which 40 are cross-service, yielding precision 0.225, recall 0.396, and lift 1.93. Thus, reviewing roughly one fifth of requests captures about two fifths of observed cross-service implementations. The selected requests do not identify a clearly heavier-workload subset: median churn is 620.5 versus 547 changed lines, and lead time 15.1 versus 19.9 days, with equal medians for commits and developers.

As a simple alternative, ranking by distinct service mentions yields project-macro Precision@20\% of 0.555 and Recall@20\% of 0.325, compared with 0.572 and 0.364 for the learned ranking; the difference is not significant ($p_{\mathrm{Holm}}=0.31$). Component/scope-label ranking yields 0.274. Hence, the learned approach does not demonstrate a reliable project-level advantage over the service-mention heuristic.

\textit{Answer to RQ4.} No individual evidence family or the first day of discussion resolves most scope uncertainty. Early predictions can nevertheless concentrate some cross-service implementations for review, although a transparent heuristic offers comparable triage performance.

\paragraph{Robustness summary.} Deduplicating 26 exact implementation-signature groups leaves 837 requests and preserves the main RQ1 effects, except that lead time remains inconclusive. Removing Camunda preserves positive effects for commits, developers, and churn, but raises cross-service prevalence to 48.6\% and changes the best RQ2 model. Cross-service prevalence also varies across traceability mechanisms (7.6\%--22.8\%). Accordingly, the workload and limited-predictability conclusions remain, whereas exact prevalence and model ranking are project- and evidence-dependent.

\section{Discussion}
\label{sec:discussion}

Our findings reveal an important boundary in what can be inferred from feature requests before implementation begins. Request-time evidence contains a modest signal about where implementation may spread, but substantially less information about how much work will follow. Cross-service requests show broader realized implementations, particularly higher code churn, although the corrected evidence does not establish longer lead time. The churn direction persists under no-Camunda and deduplication analyses. However, prevalence varies strongly across projects and model rankings change across sparse-fold and seed checks; the best model's advantage over the majority baseline is not significant after Holm correction. Later architectural and coordination decisions may also be absent from initial requests. The contribution is therefore not a universally accurate scope predictor, but evidence that cross-service scope is partially observable early while exact workload remains difficult to anticipate.

For practitioners, such predictions are best used for scope triage. A request ranked as likely cross-service can trigger earlier inspection of architectural reach, clarification of service-boundary assumptions, or involvement of developers familiar with neighboring services. Selecting roughly the top 20\% within each project prioritizes 178 requests and captures 40 of 101 cross-service implementations. Among the selected requests, 138 remain single-service, while 61 cross-service requests fall outside the selection, illustrating potential review burden and missed cases. A simple service-mention heuristic performs comparably at the project level. Importantly, prioritized requests do not form a consistently heavier-workload subset. Thus, predictions should guide attention rather than serve as effort estimates or indicators of architectural harm; legitimate features, migrations, and dependency changes may also cross service boundaries.

Several limitations qualify these implications. The cohort contains only requests with repository-recoverable issue--PR--commit traceability, and cross-service prevalence varies across traceability mechanisms. Human adjudication removed spurious late references that had distorted lead-time measurements, but residual labeling uncertainty remains. Service boundaries are derived from validated repository mappings rather than runtime deployment or formal ownership, while project sizes are highly uneven. The binary scope label also abstracts away the intent and necessity of boundary crossing. Sensitivity analyses expose, but do not eliminate, dependence on project dominance, duplicate implementations, and sparse folds.

Future work should therefore emphasize richer architectural and organizational context rather than simply more complex predictors. Industrial systems, deployment and ownership information, and finer-grained categories distinguishing feature-driven work, dependency updates, migrations, and refactoring may explain scope more effectively. Human-in-the-loop studies should further evaluate whether early scope signals improve review assignment, architectural questioning, or coordination planning. Overall, feature-request evidence appears most valuable for directing limited review attention under uncertainty, not for replacing engineering judgment or predicting downstream workload precisely.

\section{Threats to Validity}
\label{sec:threats}

We discuss threats following Wohlin et al.~\cite{wohlin2012experimentation}.

\textbf{Construct validity.} Implementation scope is measured by the services touched by linked commits, while workload is approximated through commits, developers, churn, and lead time. These repository-observable proxies do not directly capture architectural intent, ownership, coordination effort, or business cost. We therefore separate scope from workload and do not interpret cross-service implementation as inherently harmful. The retrospective triage evaluation does not measure actual review costs or benefits; moreover, service-mention features use retrospectively validated service vocabularies that may not fully reflect information available at request time.


\textbf{Internal validity.} Errors in issue--PR--commit traceability, duplicate implementation evidence, or file-to-service mapping may affect reconstructed outcomes. We mitigate these risks through repository-native links, validated service mappings, targeted human adjudication, deduplication sensitivity, and comparisons across traceability mechanisms. Adjudication corrected 69 Camunda requests, but did not exhaustively verify every link. A separate 50-case probability sample, used for validation only, estimated residual labeling error at 3.0\% (design-based 95\% CI: 0--6.7\%). The disappearance of the original lead-time association after correction illustrates sensitivity to spurious late references. Predictors are frozen at their checkpoints, and preprocessing, tuning, and fitting remain inside training folds to limit leakage.


\textbf{External validity.} The cohort contains 866 feature-labeled requests with repository-recoverable implementations, excluding requests without sufficient traceability. Project sizes and cross-service prevalence are highly uneven: Camunda contributes 720 requests, while several other projects have very small samples. LOSO validation, project-macro aggregation, no-Camunda analysis, and adequate-support sensitivity reduce dependence on individual systems, but findings may differ in industrial settings with different architectures, ownership structures, and governance practices.


\textbf{Conclusion validity.} Cross-service cases are relatively sparse, and some held-out projects provide only one class. So we report confidence intervals, effect sizes, project-macro metrics, multiplicity-corrected comparisons, and within-project robustness checks. The combined model's advantage over the majority baseline is not significant after Holm correction ($p=0.078$), and model rankings vary across seeds and sensitivity analyses. The local LLM rerun used a different build of the same model, with measured output drift. Accordingly, conclusions emphasize limited early scope signal rather than superiority of a particular model, precise workload prediction, or automated implementation decisions.

\section{Conclusion}
\label{sec:conclusion}

This paper examined how much of a feature request's downstream implementation can be anticipated before implementation begins. Across 866 adjudication-corrected, repository-traceable requests from 13 open-source microservice projects, cross-service implementations represented 11.7\% of the pooled cohort, although their prevalence varied substantially across projects. When boundary crossing occurred, it was associated with broader repository-observable work, most consistently in code churn; a longer lead time was not established. Pre-implementation evidence provided limited cross-project predictive signal: the strongest primary model reached a project-macro Macro-F1 of 0.484, but its advantage over the majority baseline was not significant after Holm correction and was sensitive to small test folds and random seeds. The local LLM provided no reliable improvement, while a simple service-mention heuristic performed comparably to learned scope triage. Exact downstream workload proved substantially harder to estimate, with learned models failing to consistently outperform training-fold median baselines. Overall, these results support lightweight cross-service scope triage rather than precise workload estimation or automated implementation planning.


\paragraph{Data Availability.} Replication code, available processed data, ground-truth validation records, and reference results are archived at \url{https://doi.org/10.5281/zenodo.23757883}. Original repository-mining data, third-party issue text, and LLM prompts are not publicly redistributed. The public package supports RQ1 reproduction and ground-truth verification; reproducing RQ2–RQ4 requires the withheld text component, subject to the access conditions described in the README.

\paragraph{Disclosure of Interests.} The authors have no competing interests to declare that are relevant to the content of this article.

\bibliographystyle{splncs04}
\bibliography{bib}
\end{document}